\documentclass[prl,preprint,showpacss]{revtex4}
\usepackage{amsmath}
\usepackage{epsfig}
\usepackage{rotating}

\begin{document}

%\titlefigure[width=5cm]{wfs.eps}
%\titlefigurecaption{
%\hbox{~}\hfill This is the Abstract-figure caption.\hfill\hbox{~}}

%%%%%%%%%%%%%%%%%%%%%%%%%%%%%%%%%%%%%%%%%%%%%%%%%%%%%%

\title{Band bending and quasi-2DEG in the 
metallized $\beta$-SiC(001) surface}

\author{R.~Rurali}
\affiliation{Departament d'Enginyeria Electr\`{o}nica,
             Universitat Aut\`{o}noma de Barcelona
             08193 Bellaterra, Spain}

\author{E.~Wachowicz}
\affiliation{Institute of Experimental Physics,
             University of Wroc\l{}aw,
             PL-50204, Wroc\l{}aw, Poland}
\affiliation{Department of Applied Physics,
             Chalmers University of Technology,
             SE-41296 G\"{o}teborg, Sweden}

\author{P.~Hyldgaard}
\affiliation{Department of Applied Physics,
             Chalmers University of Technology,
             SE-41296 G\"{o}teborg, Sweden}
\affiliation{Dept.~of~Microtechnology~and~Nanoscience, Chalmers University
             of Technology, SE-41296 G\"{o}teborg, Sweden}

\author{P. Ordej\'{o}n}
\affiliation{Centre d'Investigaci\'{o} en Nanoci\`{e}ncia
             i Nanotecnologia - CIN2 (CSIC-ICN),
             Campus UAB, E-08193 Bellaterra,
             Spain}

% \mail{e-mail
% \textsf{pablo.ordejon@cin2.es}}

%%PACS-Numbers

\begin{abstract}
We study the mechanism leading to the metallization of
the $\beta$-SiC(001) Si-rich surface induced by hydrogen
adsorption. We analyze the effects
of band bending and demonstrate the existence of a 
quasi-2D electron gas, which originates
from the donation of
electrons from adsorbed hydrogen to bulk conduction states.
We also provide a simple model that captures the 
main features of the results of first-principles calculations,
and uncovers the basic physics of the process.
\end{abstract}

\maketitle

\begin{figure}[t]
%\begin{center}
\includegraphics[width=70mm]{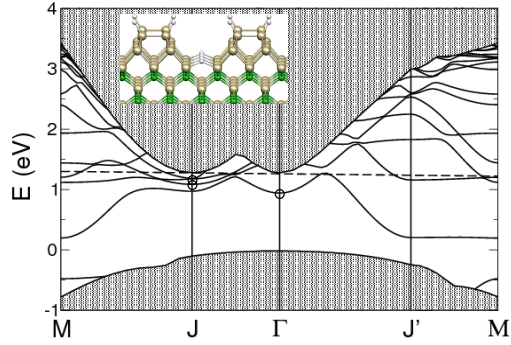}
%\end{center}
\caption{Band structure of the model of the metallized surface 
(shown schematically in the inset). The broken line is
the Fermi level. Circles at $\Gamma$ and~$J$ indicate the
quasi-2D bands.}
\label{fig:bands}
\end{figure}

The Si-rich $\beta$-SiC(001) $3 \times 2$ surface has been
reported to become metallic
when exposed to atomic hydrogen~\cite{amy,natmat}.
This is quite surprising, as H-saturation of semiconductor surfaces
normally leads to passivation.
The experimental work has shown that the metallization
does not involve the surface silicon dimers, which are
passivated upon hydrogen or water exposure~\cite{amy}.
It was early speculated that the metallization
was caused by the incorporation of hydrogen in deeper
layers, and more specifically in the
troughs  of the $3\times 2$ reconstruction.
Recent theoretical
works~\cite{difelice,chang,peng} have
shown that this is a plausible (although not
thoroughly accepted) explanation for the
surface metallization. Fig.~\ref{fig:bands} shows the
band structure of the proposed configuration 
(schematically shown in the inset),
obtained from a Density Functional Theory (DFT) calculation~\cite{method}
in a thick slab containing the top Si layers plus
12 C-Si bilayers.
The shadowed area shows the projection of the bulk
band structure, which is positioned with respect to
the surface bands by aligning the electrostatic potential
of the bulk with that of the slab calculation at
a depth where this potential becomes flat.
The surface is clearly metallic, with several band crossing
the Fermi level. 

As discussed by other authors~\cite{difelice,chang,peng},
while some of the metallic bands 
correspond to surface states,
others come from states deriving from the conduction
band of the host SiC crystal. These states can be seen 
in Fig.~\ref{fig:bands} as parabolic bands
split off the conduction band minima at the  
$J$ and $\Gamma$ points of the Surface Brilloiun Zone (SBZ), 
while the extended conduction band states remain above
the Fermi level.
In this letter, we study these interesting and quite
unexplored aspects of this surface, and 
show how these states become localized near the surface
forming quantized quasi-two dimensional states due to band bending
effects. 

\begin{figure}[b]
%\begin{center}
\includegraphics[width=70mm]{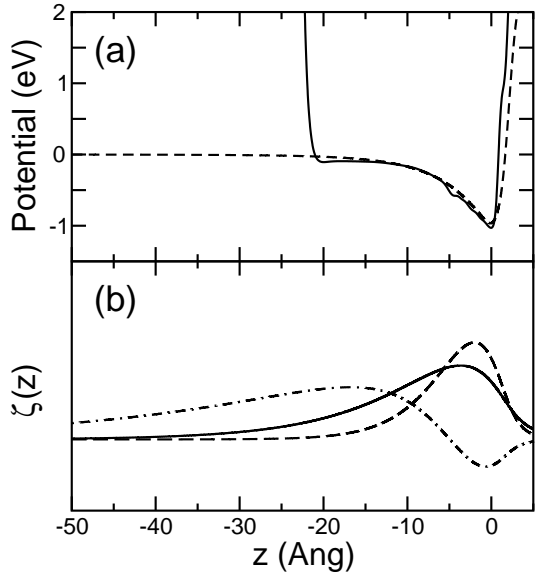}
%\end{center}
\caption{(a) Average potential seen by the electrons from the 
         DFT slab calculations
         (full line) and the effective mass model (broken line).
         (b) $z$-dependent part of the bound wave functions 
         obtained from the effective mass model.
         Dashed and dot-dashed lines: bound solutions
         for the longitudinal valleys; full line:
         bound solution for the transverse valley.
         The origin in the z axis is 
         chosen at the position of the Carbon layer closest to the surface.}
\label{fig:model}
\end{figure}

\begin{figure}[t]
%\begin{center}
\includegraphics[width=75mm]{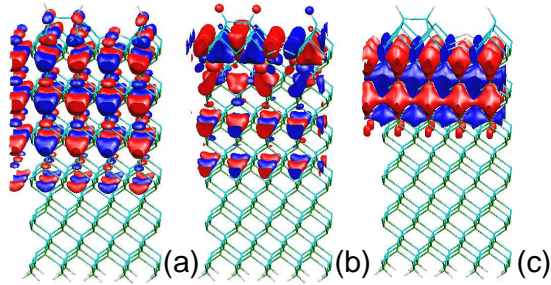}
%\end{center}
\caption{Constant contour plots of the 
DFT wave functions of the quasi-two dimensional
states at the $J$ point (panels (a) and (b)), and 
the $\Gamma$ point (panel (c)). Blue and red indicate
positive and negative values, respectively.}
\label{fig:wfs}
\end{figure}

The key to understand surface metallization is
the realization that incorporation of hydrogen
to the in-trough sites leads to transfer of electrons
from the Si-H-Si center to surface and conduction
bands that were unoccupied prior to hydrogen incorporation.
The separation
of charges leads to
electric fields which produce a strong band bending 
near the surface. 
Fig.~\ref{fig:model}(a) (full line) shows the averaged
electrostatic potential~\cite{macroave} obtained in our
DFT slab calculations as a function of the depth.
These results show the presence of electric fields at depths
of more than 10 \AA~from the Si-rich region,
well into the host SiC material.
This band bending acts as a quantum
well potential that produces
quantization of the conduction band states,
which become localized 
in the direction perpendicular to the surface. 
Fig.~\ref{fig:wfs} shows these states, as obtained
in the DFT calculations. 

To provide a deeper understanding of the band bending
and the appearance of quantized states, we 
present a simple model which 
captures the physics of the system.
We use the method proposed by Stern and Howard\cite{stern,ando}
to study semiconductor inversion layers. 
We only consider the electrons donated to the conduction band 
by the in-trough adsorbed hydrogen atoms.  The rest of the electrons
(including those donated to surface bands) are not considered
explicitly, since they do not contribute
significantly to band bending effects.
We work within the effective mass approximation, using the effective
mass tensor corresponding to the bottom of the conduction band.
Since we are dealing with the (001) surface orientation, the 
conduction band has three minima in the SBZ of the 1$\times$1 
surface reconstruction.
The minima correspond to the 
conduction band valleys at the $X$ points of the bulk $\beta$-SiC 
Brillouin zone,
giving three ellipses with longitudinal and transverse
masses of 0.68 and 0.25 $m_0$, respectively\cite{kaplan}.
Two of the ellipses have their longitudinal mass in the
surface plane, and are centered at the $M$ points
of the $\beta$-SiC(001) 1$\times$1 SBZ.
The third one has its longitudinal mass along
the direction perpendicular to the surface, 
and is centered at the $\Gamma$ point of the
$\beta$-SiC(001) 1$\times$1 SBZ.
Note that the $\Gamma$ and $M$ points 
of the 1$\times$1 reconstruction
fold down to the $\Gamma$ and $J$ 
points of the 3$\times$2 reconstruction,
which is the reason why the minima of the SiC 
bulk conduction
bands projected in the 3$\times$2 SBZ appear, 
at the same energy, at $\Gamma$ and $J$ in Fig.~\ref{fig:bands}

By the use of the effective mass approximation, 
the problem is reduced to finding the solutions
of the potential 
in the direction perpendicular
to the surface, while in the parallel
directions the solutions are plane waves with parabolic dispersion.
We model the effective potential $V_{eff}(z)$ seen by the electrons
as the contribution of two terms.
The first one represents the confining potential at the
surface, which keeps the electrons inside the bulk.
Consistently with the effective mass approximation,
this potential is taken
as zero inside the bulk and to raise smoothly at the 
surface to reach the vacuum level $V_{vac}$ outside the solid. 
Here, $V_{vac}$ is the energy required for an electron at the bottom
of the conduction band to escape the surface, which we
estimate from the difference 
between the work function of the clean surface and the 
energy gap of SiC, resulting in a value of 3.5 eV. 
The second 
term describes the electrostatic attraction between the 
conduction electrons and the net compensating positive charge 
at the surface layers, as well as the repulsion between the 
conduction electrons. The latter is described in the Hartree
approximation\cite{stern}, thus neglecting exchange and correlation
effects. 
The electrostatic term is related to the charge density 
of the conduction band electrons and the associated
positive charge left in the adsorbed hydrogen atoms by
Poisson's equation.

We solve numerically  the 1D
Schr\"odinger effective mass equations relating the potential 
$V_{eff}(z)$ and the wavefunctions
$\zeta_i^a(z)$ and their energies $E_i^a$
for each of the conduction band valleys
of the SiC(001) surface. Here, $a=l$ or $t$ indicates
the longitudinal or transversal valley, and $i$ is an index
labeling each of the bound solutions.
The bound states are occupied 
up to the Fermi level, which is determined by the number of 
electrons donated by the adsorbed hydrogen atoms. In all the calculations,
the electronic temperature was set to zero, so the
electronic occupations follow the step function 
distribution ({\em i.e.}, there are no thermally excited
carriers). Once the 
occupations of the bound states are determined, the charge 
density is constructed and Poisson's equation is solved 
numerically to obtain the 
updated effective potential $V_{eff}(z)$. This procedure is 
repeated until self-consistency is achieved.

The self-consistent electron potential obtained
from the model is shown in Fig.~\ref{fig:model}(a). It is
very similar to the DFT potential, which clearly shows that
the origin of the band bending is the donation of electrons
to the conduction band. For the quantized wave functions, we 
obtain three bound solutions. For the ellipse
perpendicular to the surface, we obtain two 
bound states, at energies 0.34 and 0.01 eV below
the bottom of the bulk conduction band, shown
in dashed and dot-dashed lines in Fig.~\ref{fig:model}(b),
respectively. This ellipse corresponds to
states at the $\Gamma$ point of the 3$\times$2 SBZ.
In the DFT results shown in Fig.~\ref{fig:bands},
only one bound state is observed, with an energy
close to that found in the model for the lowest
state. The higher one is not seen as a bound
state in the DFT calculation, a discrepancy
which is explained by 
the finite thickness of the
slab used in the DFT calculation (note that this
state extends several tens of \AA ngstr\"om
from the surface, much more than the slab
thickness). 
For the ellipses parallel to the surface plane, we
obtain only one bound state (full line in Fig.~\ref{fig:model}(b)),
with an energy 0.16 eV below the bulk conduction band, which
is doubly degenerate due to the existence of two equivalent
ellipses.
It corresponds to states in the $J$ point of the 3$\times$2
SBZ.
In the DFT calculation, due to the presence of the surface reconstruction
that breaks the crystal symmetry in the $xy$ plane (due to the
presence of the dimer rows and troughs), these two
states mix up and split. Also, they mix with 
other surface states 
at similar energies at the $J$ point point.
The resulting DFT wave functions are those shown in 
Fig.~\ref{fig:wfs}(a)-(b).
We note that the lower binding energy and 
larger spatial width 
of this state compared to the most bound state of
the ellipse perpendicular to the surface is due to the smaller
effective mass of the electrons in the $z$ direction.

Both in DFT and in the simple model, 
the underlying bulk SiC crystal is assumed to be an intrinsic
semiconductor at T=0$^o$K. Therefore, there
are no free carriers available from the bulk.
The band bending effects described here are
not due to screening of surface dipoles or charges
by bulk carriers, as it is the usual case
in semiconductor surfaces and interfaces. It is
purely due to the carriers donated by the
adsorbed hydrogen to the conduction
bands.   For substantial doping and finite temperatures,
the existence of free carriers will change
cuantitatively the profile of the band bending 
at long distances,
although the qualitative picture of band
bending originated near the surface from donation
of carriers to the conduction band remains.
For n-type doping, the bulk Fermi level will be also 
pinned at the bottom of the conduction band, so the 
picture presented here remains accurate. For p-type 
doping, an additional bending will appear at longer 
distances to align the Fermi level at the surface with 
that of the bulk (pinned around the maximum of the 
valence band). Its extension will depend on the doping 
level and the temperature, but will be
longer than the one obtained here, for reasonable values 
of these variables.

%In the case of n-type doping, the Fermi level
%of the bulk will be pinned at the bottom of
%the conduction band, so the picture presented
%here will be very accurate (besides some
%temperature-dependent extra screening
%due to bulk carriers). For p-type doping,
%there will be a band bending as the one 
%described here at distances of a few tens
%of Bohrs, plus an additional bending 
%to align the Fermi level at the surface
%(just above the bottom of the conduction band)
%with that at the bulk (pinned around the maximum
%of the valence band), whose extension will
%depend on the doping level and the temperature,
%but will typically be longer than the one
%obtained here.

Finally, it is interesting to note that the computed band
bending is able to explain the core level
shifts experimentally observed by D'Angelo {\em et al.}\cite{michel}.
Besides several surface related components, these
experiments show that the 'bulk' component of the Si $2p$ core
level spectra of the clean (unhydrogenated) surface
shifts to lower kinetic energies upon hydrogenation. The amount
of the shift depends on the H exposure, and reaches -0.84 eV 
on saturation. This 'bulk' component of the Si $2p$ core
level spectra is ascribed to the first Si layer of
the SiC underlying the three top-surface Si planes, because
due to the surface sensitivity at the photon energy used  (150.46 eV)
deeper Si layers do not contribute to the 
photoemission signal.
The core level shift of this 'bulk' component 
is therefore a measure of the 
change in the local potential at the position of the topmost
silicon layer of the SiC crystal, upon hydrogenation.
In our calculations, 
as shown in Fig.~\ref{fig:model},
this amounts to -0.93, in
excellent agreement with the experimental value.

\begin{acknowledgements}
R.~R. acknowledges the Spanish MEC (Ram\'on y Cajal Programme,
Grant No. TEC2006-13731-C02-01).
P.~H. acknowledges support from the Swedish Foundation for Strategic
Research and the Swedish National Graduate School in Materials Science.
P.~O. acknowledges
funding from Spanish MEC (Grants No. FIS2006-12117-C04-01 and CSD2007-00050), and
Generalitat de Catalunya (Grant No. SGR-2005 683).
\end{acknowledgements}

%\removelastskip

\end{document}